# Integrating 6LoWPAN Security with RPL Using The Chained Secure Mode Framework

Ahmed Raoof
Dep. of Systems and Computer Eng.
Carleton University, Canada
Email: ahmed.raoof@carleton.ca

Chung-Horng Lung
Dep. of Systems and Computer Eng.
Carleton University, Canada
Email: chlung@sce.carleton.ca

Ashraf Matrawy
School of Information Technology
Carleton University, Canada
Email: ashraf.matrawy@carleton.ca

***Abstract*—The IPv6 over Low-powered Wireless Personal Area Network (6LoWPAN) protocol was introduced to allow the transmission of Internet Protocol version 6 (IPv6) packets using the smaller-size frames of the IEEE 802.15.4 standard, which is used in many Internet of Things (IoT) networks. The primary duty of the 6LoWPAN protocol is packet fragmentation and reassembly. However, the protocol's standard currently does not include any security measures, not even authenticating the fragments' immediate sender. This lack of immediate-sender authentication opens the door for adversaries to launch several attacks on the fragmentation process, such as the buffer-reservation attacks that lead to a Denial of Service (DoS) attack and resource exhaustion of the victim nodes. This paper proposes a security integration between 6LoWPAN and the Routing Protocol for Low Power and Lossy Networks (RPL) through the Chained Secure Mode (CSM) framework as a possible solution. Since the CSM framework provides a mean of immediate-sender trust, through the use of Network Coding (NC), and an integration interface for the other protocols (or mechanisms) to use this trust to build security decisions, 6LoWPAN can use this integration to build a chain-of-trust along the fragments routing path. A proof-of-concept implementation was done in Contiki Operating System (OS), and its security and performance were evaluated against an external adversary launching a buffer-reservation attack. The results from the evaluation showed significant mitigation of the attack with almost no increase in power consumption, which presents the great potential for such integration to secure the forwarding process at the 6LoWPAN Adaptation Layer.**

## I. INTRODUCTION

The 6LoWPAN Adaptation Layer [1] and its protocol are widely used in many IoT networks to adapt the IPv6 packets to the smaller-size Link Layer frames [2]. This adaptation is responsible for the packets' fragmentation and reassembly at the nodes depending on the used forwarding scheme [1].

The security of 6LoWPAN and its common attacks was the subject of many literature [3]–[6], where it was shown that the 6LoWPAN protocol has a serious security issue: the lack of sender authentication. However, most of the proposed solutions for this problem require either extensive modifications of the 6LoWPAN protocol or using external security mechanisms that are independent of the regular uIP stack.

The Chained Secure Mode (CSM) framework [7], [8] was proposed (as a new RPL secure mode) to provide immediate-sender authentication to RPL as RPL was found to suffer from the same security issue mentioned above as in the 6LoWPAN protocol [9]. Furthermore, the CSM framework includes a trust-based integration interface (the *CSM-Trust* interface) that allows external security mechanisms or protocols to read and control the trust relationship with the node's immediate neighbors. The evaluation of CSM and its integration interface [8] showed excellent mitigation capabilities against routing replay attacks (e.g., Neighbor attack (NA) and Wormhole (WH) attacks [10]), which opens the door for more use cases, such as the one described in the following paragraph.

In this paper, a security integration between RPL and 6LoWPAN protocols is proposed using the CSM framework. Using the CSM framework and integrating a suitable trust-based external security mechanism, the 6LoWPAN protocol can use the generated *chain-of-trust*, using the intra-flow Network Coding (NC) scheme, to control fragments' admission to the assembly buffer (see §V). With a focus on mitigating the buffer-reservation attacks (see §VI-B), a security and performance evaluation is conducted on the proposed integration, which showed the great potentials for such use case, as it was able to mitigate the external adversary of the investigated attack using a simple proof-of-concept security mechanism.

The rest of the paper goes as follows: Related work is discussed in §II. A brief overview on 6LoWPAN and its common attacks is presented in §III, and another for RPL and the CSM framework is provided in §IV. Section V explains the concept behind the security integration case and its demonstration. The evaluation setup, assumptions, and adversary model are described in §VI. Section VII analyzes and discusses the evaluation results. Finally, the paper is concluded in §VIII.

## II. RELATED WORK

Hummen *et al.* in [3] provided an in-depth overview on 6LoWPAN fragmentation process and its common attacks, focusing on two types of attacks: the *Fragment-Duplication* and *Buffer-Reservation* attacks (see §III). The authors noted the lack of sender authentication problem and proposed a two-part solution: the *Content-Chaining* scheme and the *Split-Buffer* management strategy. The former uses a hashing function to *chain* the fragments of the same packet together. On the other hand, the *Split-Buffer* strategy divides the available buffer into fragment-sized *slots* to store fragments from any packet, then it uses a scoring system to make the dropping decision when a buffer overload situation occurs. The evaluation of their work showed that the *Content-Chaining* scheme successfully mitigated the fragment-duplication attack, while the *Split-Buffer* strategy reduced the effect of the buffer-reservation attacks. However, it was noted by the authors themselves and also in [4] that the Split-Buffer strategy cannot fully mitigate the buffer-reservation attacks and needs more optimization.

*SecuPAN* protocol , an intensively-modified version of the 6LoWPAN protocol, was proposed by Hossain *et al.* in [11]. The protocol uses several techniques to defend the 6LoWPAN Adaptation layer against common attacks, i.e., it introduced a nonce field to fragment headers to protect from fragments reply attacks, cryptographically-generated datagram tags (based on a Public-Private key exchange) are used to mitigate spoofing attacks, fragment duplication and fabrication attacks are defended using Message Authentication Code (MAC)-based scheme, and finally, a reputation-based buffer management strategy is used to prevent buffer-reservation attacks. The authors evaluated their implementation of the protocol against the investigated attacks, and the results showed significant reduction of the investigated attacks' effect.

## III. Overview on 6LoWPAN Adaptation Layer and its Common Attacks

The 6LoWPAN protocol [1] (and the adaptation layer) was introduced in 2007 to adapt the IPv6 packets for transmission over the smaller-sized IEEE 802.15.4 frames [12]. Two of the main functions of the protocol are the *fragmentation* and *reassembly* of IPv6 packets. This is done through several steps [1], [3]: first, compressing the IPv6 and the Transport Layer headers; secondly, dividing the packets into suitably-sized fragments; and finally, adding the associated 6LoWPAN fragmentation header to the beginning of each fragment. In 6LoWPAN, there are two types of fragmentation headers: **FRAG1**, which is used for the first fragment only, and **FRAGN** for the remaining fragments. Fig.1 shows an example of the fragmentation process as detailed in [4].

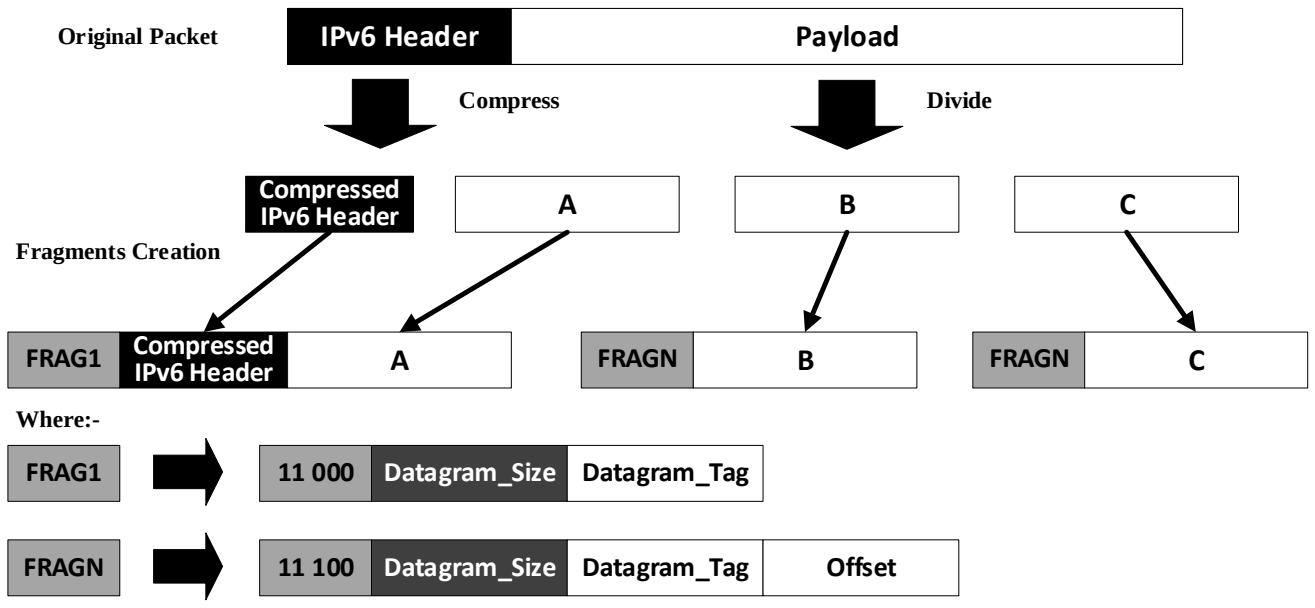

Fig. 1. An example of packet fragmentation in the 6LoWPAN protocol [4].

Sending fragments through the network does not guarantee their arrival in order. Hence, there is a need for a buffer strategy to store the fragments in the buffer until all of them are received. According to the current draft of the IoT framework [13], the 6LoWPAN protocol does not provide a specification for the buffer management, and it is left for the implementation [1]. Hence, most of the IoT OSs restrict the buffer to receive only one or two fragmented packets, due to the limited memory available on the device, e.g. Contiki OS [14] reserves enough buffer for only one fragmented packet by default. This introduces an opportunity for several types of DoS attacks to be initiated at the 6LoWPAN Adaptation Layer level, as described later on.

### *Forwarding Fragments in 6LoWPAN Networks*

According to the 6LoWPAN standard [1], [15], [16], fragments of a packet can be forwarded by intermediate nodes in one of three different ways:

1) **Route-Over**: Here, every node on the routing path must reassemble the fragmented packet and process it at the Network Layer to make a routing decision [3], [11]. Afterward, the packet is fragmented again and sent to the next hop. This method is the default option for RPL-based networks.
2) **Mesh-Under**: A 6LoWPAN mesh header is added to all the fragments that includes the *originator* and *final-destination* Link Layer addresses. The 6LoWPAN protocol will use this information and a mesh routing protocol (e.g., Ad-hoc On-demand Distance Vector (AODV) [17]) to decide the next hop. The fragments are individually forwarded and only reassembled at the destination [1].
3) **Enhanced-Route-Over**: Introduced in [15], only the first fragment (i.e., FRAG1) is sent to the Network Layer for next-hop decision, as this fragment holds the compressed IPv6 header. Then, the 6LoWPAN protocol will store the datagram tag and the next hop in a forwarding table and use it to forward all the fragments without packet reassembly or referring to the Network Layer again. If properly implemented, this method may lower data packets' latency and routing nodes' power consumption.

The work in this paper is only applicable to the Route-Over and Enhanced-Route-Over methods, as CSM can be only used when RPL is available.

### *Common Attacks on 6LoWPAN Fragmentation Process*

The 6LoWPAN protocol, as standardized in [1], does not include any security measures nor fragment authentication to protect the fragmentation process. This lack of security measures could be used to launch fragmentation attacks [3], [4]. For example, an adversary could modify or reconstruct fragmentation fields in the fragments' header, which may overwhelm receiving nodes with big amounts of uncompleted sets of fragments, causing buffer overloads and resources exhausting, and resulting in node stalling or shutdown [5].

In this paper, the focus is on one common type of fragmentation attacks [4], [5]: the *Buffer-Reservation attack*. As discussed above, when a fragment is received, most 6LoWPAN implementations at the receiving node will reserve buffer space to store only one or two fragmented packets [14], [18]. In a buffer-reservation attack, the adversary will observe the sending behavior of legitimate node(s) to estimate how frequently they send fragmented packets (by sniffing for FRAG1s). The adversary then will precede all legitimate nodes by sending a FRAG1 to the receiver and reserve the buffer for itself. Afterward, it will wait for the time-out to occur at the receiving node and sends another FRAG1 to repeat the reservation procedure, causing a DoS attack.

A more sophisticated attack could be launched by sending either several fragments (besides the first FRAG1) in each iteration of the attack, or all the fragments (of a fake payload) spread over the time-out period. However, the last attack is not as efficient as the other two methods, but it could deceive some security systems, e.g., Intrusion Detection Systems (IDSs)

## IV. Brief of RPL and The CSM Framework

RPL was standardized in 2012 as a distance-vector routing protocol for the IoT networks [19]. It builds the routing topology, called the Destination Oriented Directed Acyclic Graph (DODAG), to route the traffic toward a single *root* based on the *rank* of the nodes (a measure of the node's distance to the root), and using the Objective Function (OF), which defines the essential configurations such as the way to calculate the rank and how to choose the parents in the DODAG.

Currently, RPL has its optional security features arranged in three secure modes [10], [19]: the ***Unsecured Mode (UM)***, where RPL depends only on the Link Layer security (if available); the ***Preinstalled Secure Mode (PSM)***, where a shared preinstalled key is used to encrypt RPL's control messages and can be used for digital signature or MAC generation; and finally, the ***Authenticated Secure Mode (ASM)***, where the nodes use the preinstalled key to join the network as *leaf* nodes, then the routing nodes will obtain another key for their communications after being authenticated. For the PSM and ASM, a replay-protection mechanism (called the Consistency Checks) is available.

It was shown in [9] that RPL in PSM (with and without the replay-protection) is still prone to many attacks, with both internal[1] and external[2] adversaries, due to the lack of immediate-sender authentication. The CSM framework [7], [8] was proposed as a solution based on NC, where it demonstrated a significant mitigation of the investigated attacks.

Details on CSM framework can be found in [7], [8], but a summary is provided herein. In CSM, RPL control messages (after being prepared as per PSM procedures) are *encoded* using the concept of intra-flow NC by randomly generated integers, called the Secret Chaining (SC) values, then are sent according to RPL standard. The SC values change after every transmission, and their next values are exchanged within the sent control message using dedicated RPL Options add-ons. This process provides means of immediate-sender authentication between any node and its neighbors, which mitigates authentication-based replay attacks, e.g., NA and WH attacks.

In addition, CSM provides a trust-based integration interface, called the *CSM-Trust* interface, which allows external security mechanisms to control how RPL accept (or reject) communications from the node's neighbors. In general, the CSM-Trust interface provides access to the following values:

- ***TrustVal***: This value defines the trustworthiness of the neighbor and it is recorded per neighbor.
- ***TrustValMax*** and ***TrustValMin***: Those represent the boundary limits of TrustVal.
- ***TrustTrig***: The threshold value for the TrustVal that, if it went below it, will have RPL drop any control messages received from the neighbor.

A conceptual diagram of the CSM-Trust interface is shown as part of Fig.2.

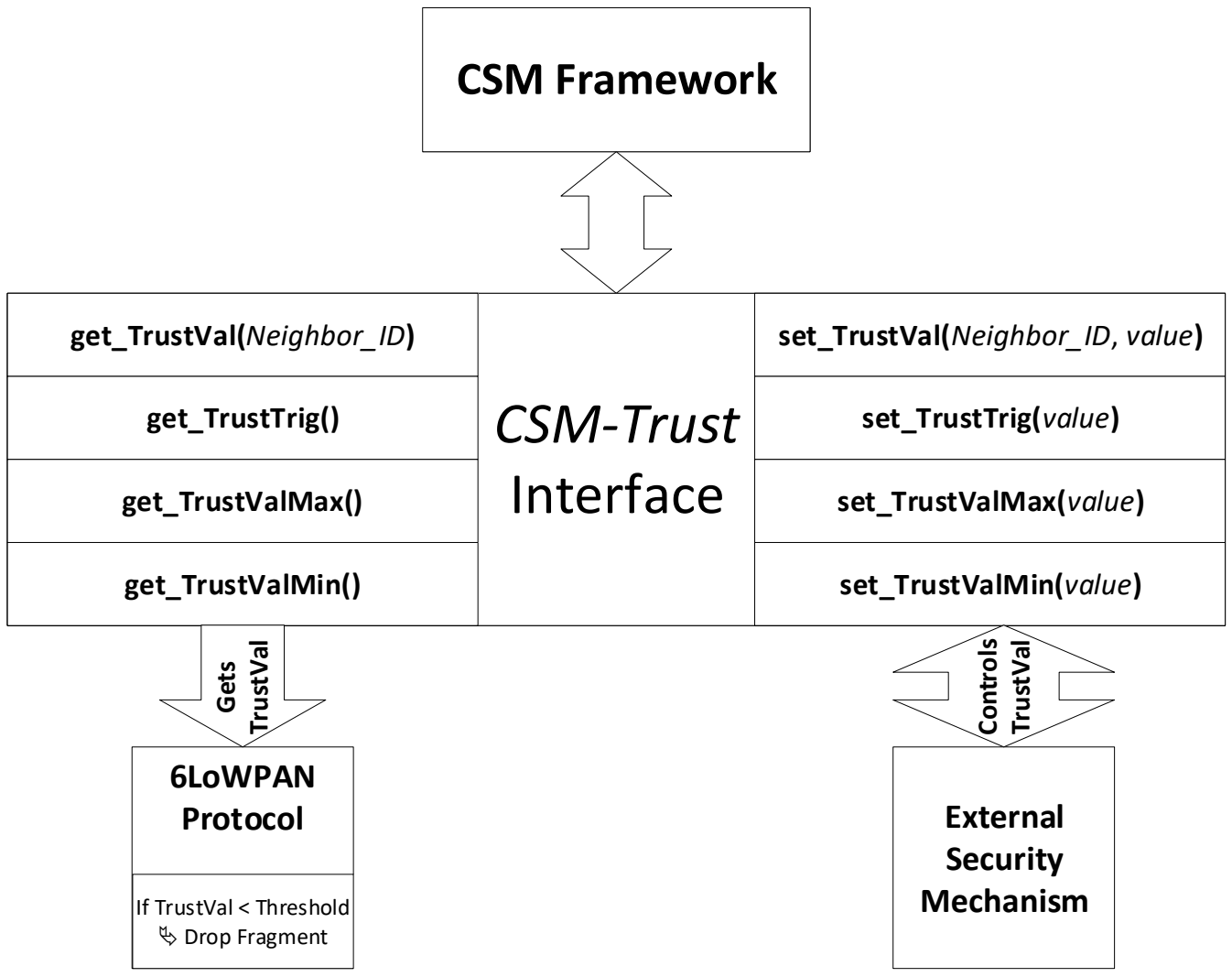

Fig. 2. CSM-6LoWPAN integration concept diagram.

## V. Integrating 6LoWPAN Security with RPL: The Concept

As mentioned in §IV, the CSM-Trust interface of the CSM framework provides a way for cross-protocol and cross-mechanism integration. An external mechanism can use its methods to define the trustworthiness of the node's neighbors and set the *TrustVal* value accordingly; hence, controlling RPL's acceptance of control messages from the node's immediate neighbors. In general, the *TrustVal* is used by RPL in CSM to secure the control plane of the Network Layer.

On the other hand, The same interface can also be used, by other protocols/mechanisms, to read the *TrustVal* value and employ it for their decision-making. This capability forms the basis of the work in this paper: using the *CSM-Trust* interface to control the fragments traffic at the 6LoWPAN Adaptation Layer, i.e., the data plane of the Network Layer.

The general concept for the security integration case (see Fig.2) is that if each node trusts its neighbors through the CSM framework at the control plane, then the nodes on the routing path for a fragmented data packet (at the data plane) can also be trusted due to the *chain-of-trust*. In other words, if the first routing node trusts the original sender of the fragments, and the second routing node trusts the first routing node, and so on for the remaining routing nodes until the destination node, then the whole path is considered secure.

[1] An internal adversary is an adversary who is part of the network, e.g., has the encryption keys used by the legitimate nodes for RPL in PSM or ASM.

[2] An external adversary refers to an adversary who is not part of the network, e.g., it does not have the encryption keys used by the legitimate nodes for RPL in PSM/ASM, or runs RPL in UM.

### A. The Security Integration Case Implementation

The goal of our proposed security integration case is to mitigate the buffer-reservation attacks originating from external adversaries on the 6LoWPAN Adaptation Layer.

As a demonstration for the security integration case, the following simple proof-of-concept mechanism was implemented:

- *TrustValMin*, *TrustValMax*, and *TrustTrig* were set to 0, 100, and 50, respectively.
- For the first RPL message from a neighbor, a successful reception will set *TrustVal* to *TrustValMax*.
- Afterward, TrustVal will increase or decrease based on the successful (or unsuccessful) decoding of the received RPL control messages. The increment (and decrement) amount was set to 10.

This external security mechanism will control the *TrustVal* value, while the 6LoWPAN protocol will use the *TrustVal* value to decide on admitting the received fragment to the assembly buffer or not, all by using the *CSM-Trust* interface of the CSM framework.

Other worth-mentioning implementation points are:

- The decisions of the 6LoWPAN protocol are based on the trust of the immediate sender of the fragment, not the original sender, fulfilling the concept of *chain-of-trust*. The protocol will extract the sender's Link Layer address from the frame containing the fragment and find the associated IPv6 address. Then, it will use the found IPv6 address to get the corresponding TrustVal using the *CSM-Trust* interface.
- Similar to CSM, the 6LoWPAN protocol will have its own threshold to make its decisions, the (6LOWPAN_TRUST_THRESHOLD). For demonstration purposes, it was set to 60.
- Currently, the Contiki OS implementation of 6LoWPAN supports only the Route-Over method [14].

## VI. Evaluation of The Security Integration Case

To evaluate the proposed security integration case, a comparison of security and performance was conducted between the vanilla 6LoWPAN protocol (as in the Contiki OS implementation) and the 6LoWPAN with CSM framework integration (CSM-6LoWPAN for short). Both protocols were tested against an external adversary of the buffer-reservation attack in several scenarios.

### A. Evaluation Setup and Assumptions

Cooja [20], the simulator for Contiki OS, was used for all the simulations (with simulated motes). Fig.3 shows the topology used in the evaluation, which is widely used for 6LoWPAN evaluations [3]–[5], [11]. A list of simulation parameters is provided in Table I.

For the evaluation, two metrics were used: the average data packet delivery rate (PDR) and the average power consumption for the legitimate sending node (per received data packet).

The following assumptions were also considered in the evaluation: both the legitimate node and the adversary sends

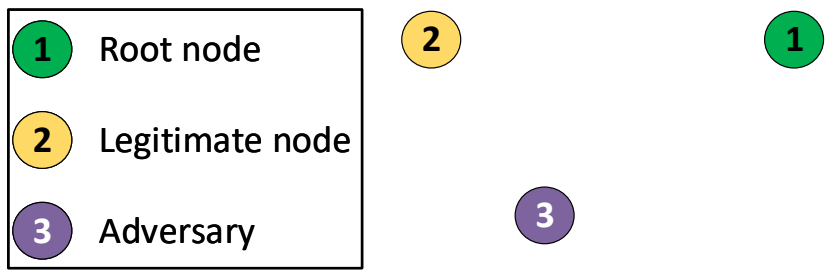

Fig. 3. Network topology used for the evaluation.

TABLE I
List of Simulation Parameters

| **Description** | **Value** |
|---|---|
| No. of experiments | Two: vanilla 6LoWPAN and CSM-6LoWPAN |
| No. of scenarios per experiment | Ten - See §VI-B |
| Sim. rounds per scenario / time | 10 rounds / 20 min. per round |
| Sensor nodes type | Arago Sys. Wismote mote |

(512 bytes) data packets toward the root at a rate of 1 packet/minute per node. However, the adversary follows the attack scenarios as described in §VI-B.

For the legitimate node, RPL is set to operate in either UM (vanilla 6LoWPAN experiment) or CSM (CSM-6LoWPAN experiment). In both cases, Contiki OS is using the default settings for its uIP stack: IEEE 802.15.4 [12] for the Physical Layer and Medium Access Control (MAC) sub-layer, ContikiMAC [21] for the Radio Duty-Cycle (RDC) sub-layer, IPv6, 6LoWPAN, and RPL at the Network Layer, and User Datagram Protocol (UDP) for the Transport Layer. In addition, all security measures and encryption at the Link Layer were assumed to be disabled.

For the 6LoWPAN protocol, the default Contiki reassembly timeout was used (20 seconds) and the max size for the fragment payload is set to 102 bytes (as per the standard [1]). In addition, fragment forwarding follows the Route-Over approach, as it is the only available option in Contiki OS.

Finally, the simulations results were averaged over ten rounds per experiment with a 95% confidence level.

### B. Adversary Model and Attack Scenarios

The adversary runs RPL in the same secure mode as the legitimate nodes. However, as an external adversary, it does not have the required encryption keys (for the CSM-6LoWPAN experiment). It is worth mentioning that there are no differences between an external and internal adversary from the UM point of view (the vanilla 6LoWPAN experiment).

In all cases, the adversary starts as a legitimate node, tries to join the network, then launches the attack after 50 seconds. This is to allow the network to reach the steady-state situation.

For each experiment, nine attack scenarios were used to simulate the different cases an adversary can initiate the buffer-reservation attack, in addition to one *No Attack* scenario for each experiment. These scenarios are summarized in Table II.

It is worth mentioning that the adversary and the legitimate node send their fragments at their designated times ± 2 seconds, due to the randomness nature of the simulation, which mimics the real-life situation.

TABLE II
SUMMARY OF THE SIMULATION SCENARIOS

| | **Adversary sends...** | | | |
|---|---|---|---|---|
| | Full Packets (Normal DoS) | 1st Fragment Only (Basic Buffer-Resrv. Attack) | All Except Last Fragment (Soph. Buffer-Resrv. Attack) | |
| **Attack Launch Timing** | Before Legitimate Node | Before Legitimate Node | Before Legitimate Node | No Attack Scenario |
| | Simultaneously with Legitimate Node | Simultaneously with Legitimate Node | Simultaneously with Legitimate Node | |
| | After Legitimate Node | After Legitimate Node | After Legitimate Node | |

## VII. Results Analysis and Discussion

### A. Results Analysis

Looking at Fig.4a for the PDR results, it is clear that CSM-6LoWPAN mitigated the attack successfully in all the different scenarios (PDR $\approx$100%). For the vanilla 6LoWPAN, the attack was more successful when the adversary preceded the legitimate node (PDR $\approx$10–60%) as the adversary reserved the buffer before the legitimate node. In addition, sending one or a few fragments was slightly more effective than sending full packets (PDR $\approx$60–75%, compared to 80–85%), except for the case of the adversary sending fragments before the legitimate node, where the attack was more successful when the adversary sent full packets (PDR $\approx$13%). However, the reason for the later case seems to be due to the simulation random irregularities; as the logs of the ten rounds suggests that the adversary always preceded the legitimate node by enough time to guarantee the buffer reservation.

Moving to the average power consumption readings in Fig.4b, it can be seen that CSM-6LoWPAN did not introduce any additional power consumption at the legitimate node compared to the vanilla 6LoWPAN experiment (around 0.1 milliwatt for CSM-6LoWPAN, compared to 0.15–0.20 for vanilla 6LoWPAN). All while providing better security (mitigating the buffer-reservation attack from an external adversary). However, it is worth mentioning that power consumption depends on the routing topology, the complexity of the used external security mechanism, and the links quality.

### B. Discussions

The observations from the evaluation experiments can be summarized in the following:

- In this paper's demonstration, the integration of the simple external security mechanism with the CSM framework (see §IV) was able to mitigate the external adversaries of the simple buffer-reservation attacks at 6LoWPAN Adaptation Layer.
  - However, since the mechanism does not have global view of the network or consider the behavior patterns of the neighbors, it doesn't provide mitigation capability against the internal adversaries of the buffer-reservation attacks.
  - In addition, a more sophisticated adversary can still launch a buffer-reservation attack if the adversary used the Link-Layer address of the victim node(s).
- The work in this paper demonstrated some possibilities that the CSM framework can brings to IoT networks through the *CSM-Trust* interface, allowing integration not only between one external security measure and RPL, but among several security measures at once.
- The security integration case also shows that the lack of immediate-sender authentication in 6LoWPAN protocol can be mitigated through the integration with the CSM framework, without heavily taxing the limited resources of the IoT devices, unlike many of the currently proposed solution [5], [6], [11].

## VIII. Conclusion

In this paper, a solution to the lack of immediate-sender authentication was proposed as a security integration case of the CSM framework. The security integration between the 6LoWPAN, RPL, and an external security mechanism, through the *CSM-Trust* interface, could provide better security to the whole IoT network due to the chain-of-trust provided by the use of NC in CSM. A proof-of-concept demonstration of the security integration case, using a simple external routing-security mechanism integrated with RPL using the CSM-Trust interface, was implemented in Contiki OS. The evaluation of this security integration case showed the potentials of the CSM framework, as the simple security integration between the three components (6LoWPAN protocol, RPL, and the external routing-security mechanism) was able to mitigate the external adversaries of a simple buffer-reservation attacks at 6LoWPAN Adaptation Layer.

Finally, this security integration case also shows that, depending on the integrated security mechanism and using CSM framework, a solution to 6LoWPAN's lack of authentication is possible without implementing resource-exhausting techniques, such as the public-private keys or new protocols.

## Acknowledgment

The authors acknowledge support from the Natural Sciences and Engineering Research Council of Canada (NSERC) through the Discovery Grant program.

## References

[1] G. Montenegro *et al.*, "Transmission of IPv6 Packets over IEEE 802.15.4 Networks," RFC 4944, September 2007. [Online]. Available: https://www.rfc-editor.org/info/rfc4944

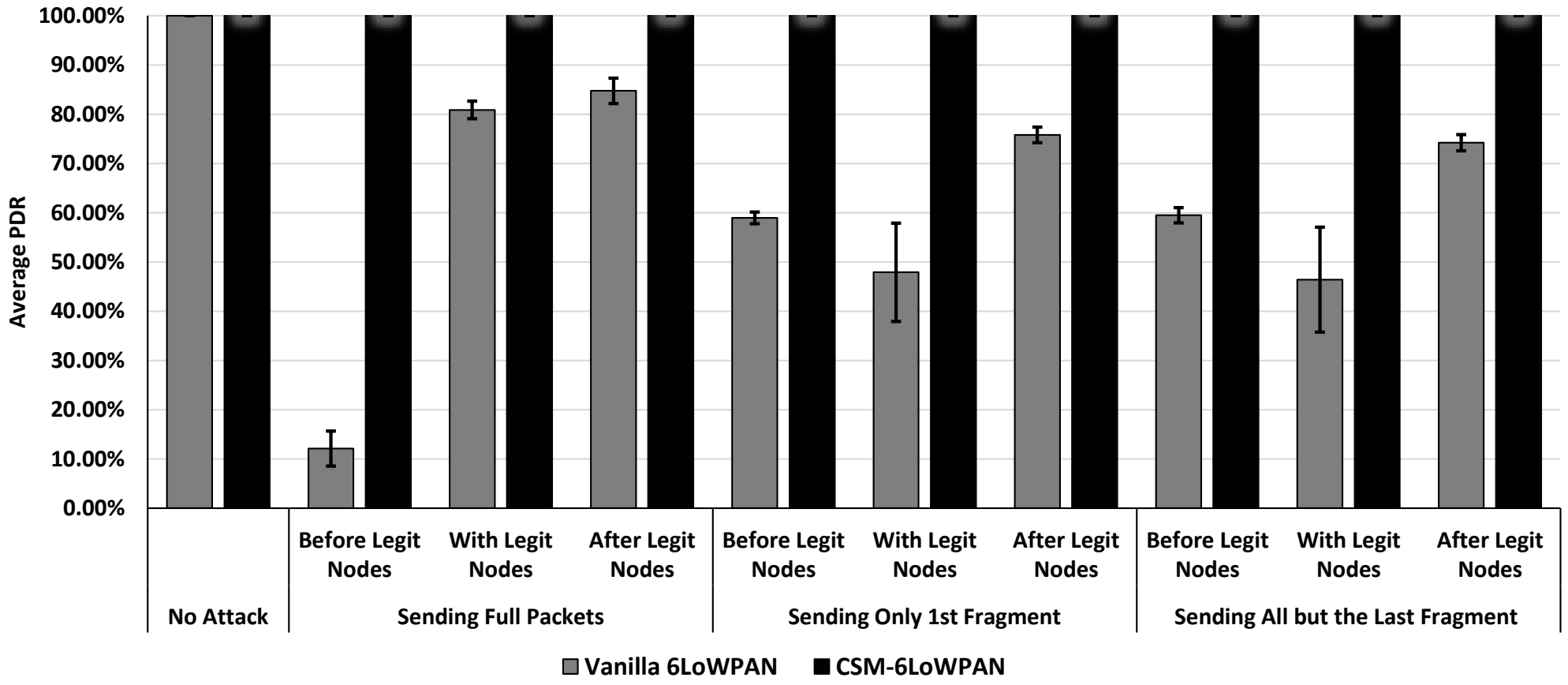

(a) Average Data Packet Delivery rate (PDR). The less the PDR, the more successful the attack.

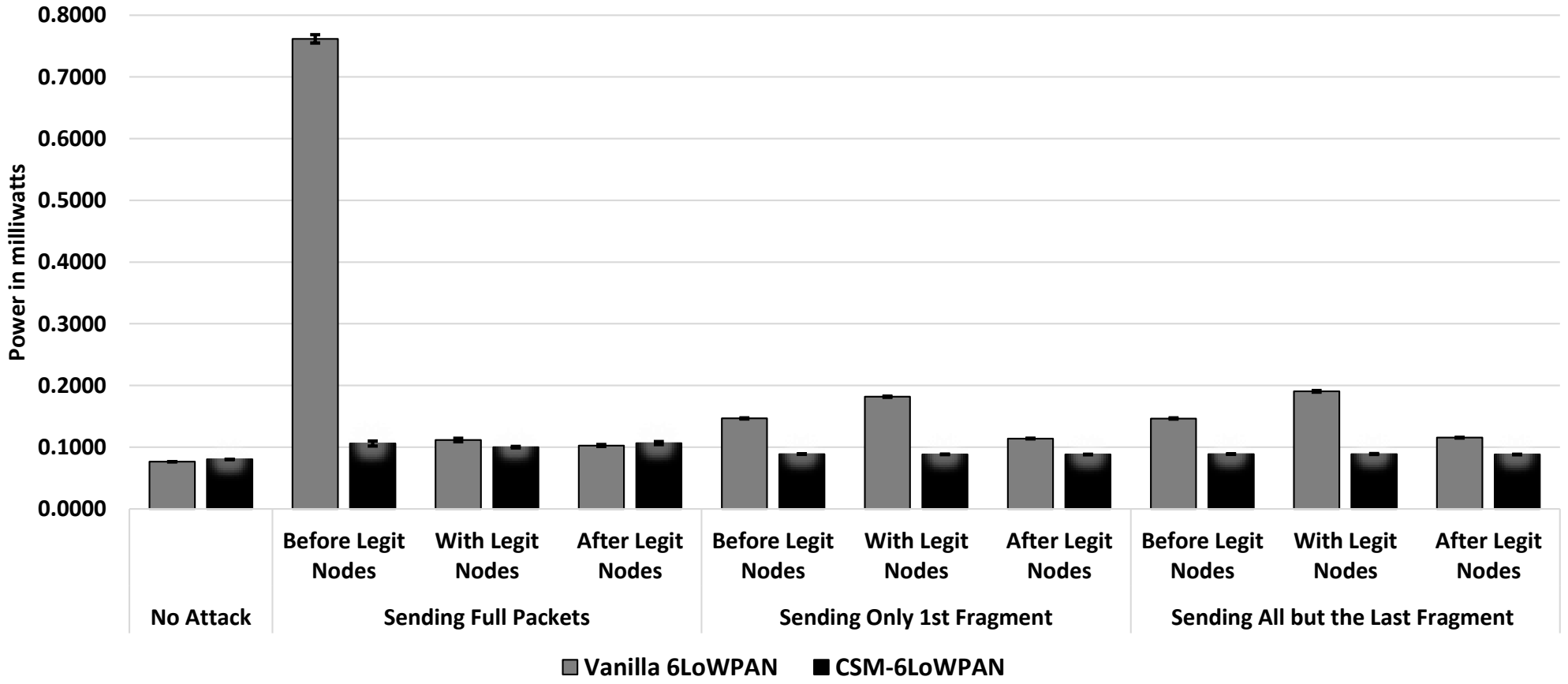

(b) Average power consumption for the legitimate sending node (per received data packet). Readings are in milliwatts

Fig. 4. Simulation results for the two experiments (vanilla 6LoWPAN and CSM-6LoWPAN.)

[2] J. Nieminen *et al.*, "IPv6 over BLUETOOTH(R) Low Energy," RFC 7668, Oct 2015. [Online]. Available: http://www.rfc-editor.org/info/rfc7668

[3] R. Hummen *et al.*, "6LoWPAN Fragmentation Attacks and Mitigation Mechanisms," in *Proc. of the 6th ACM Conf. on Security and Privacy in Wireless and Mobile Networks (WiSec 2013)*. ACM Press, 2013.

[4] A. Raoof and A. Matrawy, "The Effect of Buffer Management Strategies on 6LoWPAN's Response to Buffer Reservation Attacks," in *2017 IEEE Int'l Conf. on Communications (ICC)*, May 2017.

[5] A. K. Bediya and R. Kumar, "Real Time DDoS Intrusion Detection and Monitoring Framework in 6LoWPAN for Internet of Things," in *IEEE Int'l Conf. on Comp., Power and Commun. Tech. (GUCON)*. IEEE, 2020.

[6] G. Glissa and A. Meddeb, "6LowPSec: An End-to-End Security Protocol for 6LoWPAN," *Ad Hoc Networks*, vol. 82, Jan 2019.

[7] A. Raoof, C.-H. Lung, and A. Matrawy, "Introducing Network Coding to RPL: The Chained Secure Mode (CSM)," in *The 19th IEEE Int. Symp. on Netw. Comput. and App. (NCA)*. IEEE, 2020.

[8] ——, "Securing RPL using Network Coding: The Chained Secure Mode (CSM)," 2021, available at arXiv:2102.06254 [cs.NI].

[9] A. Raoof, A. Matrawy, and C.-H. Lung, "Enhancing Routing Security in IoT: Performance Evaluation of RPL Secure Mode under Attacks," *IEEE Internet of Things Journal*, vol. 7, no. 12, Dec 2020.

[10] ——, "Routing Attacks and Mitigation Methods for RPL-Based Internet of Things," *IEEE Comm. Surveys and Tutorials*, vol. 21, no. 2, 2nd Quarter 2019.

[11] M. Hossain *et al.*, "SecuPAN: A Security Scheme to Mitigate Fragmentation-Based Network Attacks in 6LoWPAN," in *the 8th ACM Conf. on Data and App. Secur. and Privacy (CODASPY)*. ACM, 2018.

[12] *IEEE Standard for Low-Rate Wireless Networks (802.15.4-2015)*, IEEE, April 2015.

[13] R. Minerva, A. Biru, and D. Rotondi, "Towards a definition of the Internet of Things (IoT)," May 2015. [Online]. Available: https://iot.ieee.org/definition.html

[14] A. Dunkels *et al.*, "Contiki - a Lightweight and Flexible Operating System for Tiny Networked Sensors," in *29th IEEE Int'l Conf' on Local Computer Networks*, Nov 2004.

[15] T. Watteyne *et al.*, "On Forwarding 6LoWPAN Fragments over a Multi-Hop IPv6 Network," RFC 8930, November 2020. [Online]. Available: https://rfc-editor.org/rfc/rfc8930.txt

[16] P. Thubert and J. Hui, "Compression Format for IPv6 Datagrams over IEEE 802.15.4-Based Networks," RFC 6282, September 2011. [Online]. Available: https://www.rfc-editor.org/info/rfc6282

[17] S. R. Das *et al.*, "Ad hoc On-Demand Distance Vector (AODV) Routing," RFC 3561, July 2003. [Online]. Available: https://rfc-editor.org/rfc/rfc3561.txt

[18] "RIOT: 6LoWPAN Fragmentation," RIOT OS.org, 2013, accessed on Feb $15^{th}$, 2021. [Online]. Available: https://doc.riot-os.org/group__net__sixlowpan.html

[19] T. Winter *et al.*, "RPL: IPv6 Routing Protocol for Low Power and Lossy Networks," RFC Editor, RFC 6550, March 2012.

[20] F. Osterlind *et al.*, "Cross-Level Sensor Network Simulation with COOJA," in *Proc. of 31st IEEE Conference on Local Computer Networks*. IEEE, 2006, pp. 641–648.

[21] A. Dunkels, "The ContikiMAC Radio Duty Cycling Protocol," Swedish Institute of Computer Science, Technical Report, Dec 2011.